\documentclass[useAMS,usenatbib]{mn2e}
\usepackage{url,times,graphicx,amsmath,amsfonts,amssymb,aas_macros,color,epsfig,epstopdf}
 \usepackage{xcolor}

\def \etal {et~al.~}

\newcommand{\hMpc}{{\ifmmode{h^{-1}{\rm Mpc}}\else{$h^{-1}$Mpc}\fi}}
\newcommand{\hkpc}{{\ifmmode{h^{-1}{\rm kpc}}\else{$h^{-1}$kpc}\fi}}
\newcommand{\kpc}{{\ifmmode{ {\rm kpc} }\else{{\rm kpc}}\fi}}
\newcommand{\kms}{{\ifmmode{ {\rm km\,s^{-1}} }\else{ ${\rm km\,s^{-1}}$ }\fi}}
\newcommand{\hMsun}{{\ifmmode{h^{-1}{\rm {M_{\odot}}}}\else{$h^{-1}{\rm{M_{\odot}}}$}\fi}}
\newcommand{\Msun}{{\ifmmode{{\rm M}_{\odot}}\else{${\rm M}_{\odot}$}\fi}}
\newcommand{\Mhalo}{{\ifmmode{M_{\rm halo}}\else{$M_{\rm halo}$}\fi}}
\newcommand{\Rvir}{{\ifmmode{R_{\rm vir}}\else{$R_{\rm vir}$}\fi}}
\newcommand{\Mstar}{{\ifmmode{M_{\rm star}}\else{$M_{\rm star}$}\fi}}
\newcommand{\Vrot}{{\ifmmode{V_{\rm rot}}\else{$V_{\rm rot}$}\fi}}
\newcommand{\ltsima}{$\; \buildrel < \over \sim \;$}
\newcommand{\gtsima}{$\; \buildrel > \over \sim \;$}
\newcommand{\lsim}{\lower.5ex\hbox{\ltsima}}
\newcommand{\gsim}{\lower.5ex\hbox{\gtsima}}

\def\lesssim{\mathrel{\hbox{\rlap{\hbox{\lower4pt\hbox{$\sim$}}}\hbox{$<$}}}}
\def\gtrsim{\mathrel{\hbox{\rlap{\hbox{\lower4pt\hbox{$\sim$}}}\hbox{$>$}}}}

\newcommand{\beq}{\begin{equation}}
\newcommand{\eeq}{\end{equation}}
\def\beqa{\begin{eqnarray}}
\def\eeqa{\end{eqnarray}}
\def\LCDM{\ensuremath{\Lambda}CDM}

\def\head{ \vbox to 0pt{\vss \hbox to 0pt{\hskip 440pt\rm
      LA-UR-10-07069\hss} \vskip 25pt}}

\def \kms {\ifmmode  \,\rm km\,s^{-1} \else $\,\rm km\,s^{-1}  $ \fi }
\def \kpc {\ifmmode  {\rm kpc}  \else ${\rm  kpc}$ \fi  }  
\def \hkpc {\ifmmode  {h^{-1}\rm kpc}  \else ${h^{-1}\rm kpc}$ \fi  }  
\def \hMpc {\ifmmode  {h^{-1}\rm Mpc}  \else ${h^{-1}\rm Mpc}$ \fi  }  
\def \Mpch {\ifmmode  {h^{-1}\rm Mpc}  \else ${h^{-1}\rm Mpc}$ \fi  }  
\def \Msun {\ifmmode {\rm M}_{\odot} \else ${\rm M}_{\odot}$ \fi} 
\def \hMsun {\ifmmode h^{-1}\,\rm M_{\odot} \else $h^{-1}\,\rm M_{\odot}$ \fi}

\def \LCDM {\ifmmode \Lambda{\rm CDM} \else $\Lambda{\rm CDM}$ \fi}
\def \sig8 {\ifmmode \sigma_8 \else $\sigma_8$ \fi} 
\def \OmegaM {\ifmmode \Omega_{\rm m} \else $\Omega_{\rm m}$ \fi} 
\def \Omegab {\ifmmode \Omega_{\rm b} \else $\Omega_{\rm b}$ \fi} 
\def \OmegaL {\ifmmode \Omega_{\rm \Lambda} \else $\Omega_{\rm \Lambda}$\fi} 
\def \Deltavir {\ifmmode \Delta_{\rm vir} \else $\Delta_{\rm vir}$ \fi}
\def \rhocrit {\ifmmode \rho_{\rm crit} \else $\rho_{\rm crit}$ \fi}
\def \rhou {\ifmmode \rho_{\rm u} \else $\rho_{\rm u}$ \fi}
\def \zc {\ifmmode z_{\rm c} \else $z_{\rm c}$ \fi}

\def\head{ .ps \vbox to 0pt{\vss \hbox to 0pt{\hskip 440pt\rm
      LA-UR-10-07069\hss} \vskip 25pt}} 

\def \spose#1{\hbox  to 0pt{#1\hss}}  
\def \lta{\mathrel{\spose{\lower 3pt\hbox{$\sim$}}\raise 2.0pt\hbox{$<$}}}
\def \gta{\mathrel{\spose{\lower 3pt\hbox{$\sim$}}\raise 2.0pt\hbox{$>$}}}

\title[Core creation and destruction in galaxies]
{NIHAO IV: Core creation and destruction in dark matter
  density profiles across cosmic time}

\author[E. Tollet \etal] {Edouard Tollet$^{1,2}$\thanks{E-mail:
    edouard.tollet@obspm.fr}, Andrea V. Macci\`o$^{3,1}$\thanks{E-mail:
    maccio@nyu.edu}, Aaron A. Dutton$^{3,1}$, Greg S. Stinson$^{1}$, 
  \newauthor Liang Wang$^{1,4}$, Camilla Penzo$^1$, Thales A. Gutcke$^1$,
  Tobias Buck$^1$, Xi Kang$^4$,
  \newauthor Chris Brook$^5$, Arianna Di Cintio$^6$, Ben W. Keller$^7$,
  James Wadsley$^7$\\
$^1$Max-Planck-Institut f\"ur Astronomie, K\"onigstuhl 17, 69117 Heidelberg, Germany\\
$^2$GEPI, Observatoire de Paris, CNRS,
Univ Paris Diderot, 61 Avenue de l'Observatoire,
75014 Paris, France\\
$^3$New York University Abu Dhabi, PO Box 129188, Abu Dhabi, UAE\\
$^4$Purple Mountain Observatory, the Partner Group of MPI f\"ur Astronomie, 2 West Beijing
Road, Nanjing 210008, China\\
$^5$Ramon y Cajal Fellow, Departamento de F\'isica Te\'orica, Universidad Autonoma de Madrid,
28049 Cantoblanco, Madrid, Spain\\
$^6$Dark Fellow, Dark Cosmology Centre, NBI, University of Copenhagen, Juliane Maries Vej 30, DK-2100 Copenhagen, Denmark\\
$^7$Department of Physics and Astronomy, McMaster University, Hamilton, Ontario L8S 4M1, Canada}

\begin{document}

\date{Accepted 2015 December 03. Received 2015 November 04; in original form 2015 July 13}

\pagerange{\pageref{firstpage}--\pageref{lastpage}} \pubyear{2015}

\maketitle

\label{firstpage}


\begin{abstract}

  We use the NIHAO (Numerical Investigation of Hundred Astrophysical Objects)
  cosmological simulations to investigate the effects of baryonic
  physics on the time evolution of Dark Matter central density
  profiles.   The sample is made of $\approx 70$ independent high resolution
  hydrodynamical simulations of galaxy formation and covers a wide
  mass range: $10^{10} \lta \Mhalo / \Msun \lta 10^{12}$, i.e., from
  dwarfs to L$^\star$.
  We confirm previous results on the dependence of the inner dark
  matter density slope, $\alpha$, on the ratio between stellar-to-halo mass,
  $\Mstar / \Mhalo$.
  We show that this relation holds approximately
  at all redshifts (with an intrinsic scatter of $\sim 0.18$ in $\alpha$
  measured between $1-2\%$ of the virial radius).
  This implies that in practically all haloes the shape of their inner
  density profile changes quite substantially over cosmic
  time, as they grow in stellar and total mass. 
  Thus, depending on their final $\Mstar / \Mhalo$ ratio, haloes can
  either form and keep a substantial density core ($R_{\rm core}\sim
  1$ kpc),
  or form and then destroy the core and re-contract the halo,
  going back to a cuspy profile, which is even steeper than CDM
  predictions for massive galaxies ($10^{12} \Msun$).
  We show that results from the NIHAO suite are in good
  agreement with recent observational measurements of $\alpha$ in
  dwarf galaxies.
  Overall our results suggest that the notion of a universal
  density profile for dark matter haloes is no longer valid in the
  presence of galaxy formation.

\end{abstract}

\noindent
\begin{keywords}

  cosmology: dark matter galaxies: evolution - formation -
  hydrodynamics methods:N-body simulation

 \end{keywords}

\section{Introduction} \label{sec:introduction}

\begin{table*}
\label{tab:gal}
\begin{minipage}{180mm}
\begin{center}
\caption{Properties of our three selected galaxies at z = 0}
\begin{tabular}{lllllllllllllll}
  \hline
  simulation ID & \# particles & \# DM particles & \# star particles & DM particle mass & virial mass & stellar mass & $z=0$ SFR & Conv. rad.\footnote{Following Power \etal 2003, see text for details.} \\
              & $n_{\rm 200}$ & $n_{\rm DM}$ & $n_{\rm star}$ & $m_{\rm DM}$ &($M_{200}$/M$_\odot$) &($\Mstar$/M$_\odot$)  & (M$_\odot$ yr$^{-1}$) & (kpc) \\
\hline
g2.63e10 & 458,723 & 414,291 & 18,388 &   $1.173 \times 10^4$    & $2.70\times10^{10}$ & $4.28\times10^{7}$ & 0.00 & 0.41\\

g2.19e11 & 920,447 & 557,247 & 113,958 & $2.169 \times 10^5$ &  $1.31\times10^{11}$ & $9.27\times10^{8}$ & $8.2\times10^{-2}$ & 0.65\\ 
g8.06e11 & 1,366,038 & 481,349 & 665,796 & $1.735 \times 10^6$ & $9.43\times10^{11}$ & $4.48\times10^{10}$ & 11.31 & 1.23 \\ 
\hline  
\end{tabular}\\
\end{center}
\end{minipage}
\end{table*}

The Cold Dark Matter (CDM) theory is very successful at describing the
Universe's topology and evolution on large scales (e.g. Springel \etal
2005).  This theory makes clear predictions of the distribution of
Dark Matter on small scales, where all collapsed structures (haloes)
are supposed to share an approximately self-similar dark matter
density profile, as first pointed out by Navarro, Frenk and White
(1997, NFW) and then confirmed via higher resolution numerical simulations
by a large variety of works (e.g., Klypin \etal 2001, Diemand \etal
2004; Power \etal 2003, Macci\`o et al 2007, 2008, Neto \etal 2007,
Navarro \etal 2010, Prada \etal 2012, Dutton \& Macci\`o 2014).

This prediction of a universal, cuspy profile for the dark matter
seems to be in tension with observations of rotation curves of low
mass galaxies (e.g Moore 1994, Salucci \& Burkert 2000; Simon \etal
2005; de Blok \etal 2008; Kuzio de Naray, McGaugh \& de Blok 2008;
Kuzio de Naray, McGaugh \& Mihos 2009; Oh \etal 2011a), which seem to
prefer density profiles with a constant density core in the center.

This cusp-core controversy suggests that either a modification of the
whole CDM paradigm is required (e.g., self interacting dark matter,
SIDM, Vogelsberger \etal 2014), or the inadequacy of pure N-body
simulations to capture the dark matter dynamics on small scales, due to
the absence of dissipative phenomena connected to baryonic physics.

Recent and more accurate cosmological hydrodynamical simulations have
indeed shown that baryons are able to alter the dark matter profile
and to create substantial cores (up to $\sim 1$ kpc) in the dark matter
distribution (e.g. Governato \etal 2010, Macci\`o \etal 2012, Martizzi
\etal 2013, Munshi \etal 2013, Di Cintio 2014a,b, Trujillo-Gomez \etal
2015, O{\~n}orbe  \etal 2015).

As nicely  described in Pontzen  \& Governato (2012) the  formation of
dark  matter  cores  is  linked  to the  rapid  change  of  the  total
gravitational  potential.  In the  central  kiloparsecs the  potential changes  on
sub-dynamical time-scales, as
repeated,  energetic feedback  generates large  underdense bubbles  of
expanding   gas.
The fluctuations  in the  central
potential irreversibly  (e.g. non adiabatically) transfer energy into
collisionless particles, thus generating a dark matter core
(see also Ogiya \& Mori 2014 for a detailed analysis of the effect of resonances between DM particles and the density wave excited by the oscillating potential).
These strong gas outflows are triggered by
stellar (and AGN) feedback (Navarro, Eke \& Frenk 1996; Read \&
Gilmore 2005; Mashehenko, Couchman \& Wadsley 2006; Pontzen \&
Governato 2014; Martizzi \etal 2013).

Recently Di Cintio \etal (2014a, DC14 hereafter) has shown that the
modification of the initial dark matter profile (either leading to an
expansion or a contraction) is clearly linked to the {\it integrated
  star formation efficiency} of the galaxy, which can be parameterized
through the redshift zero stellar mass - halo mass ratio
($\Mstar/\Mhalo$), and that this holds for a large variety of stellar
feedback implementations.

They clearly showed that the halo response to star formation is non
monotonic and that haloes with very low and very high star formation
efficiency (or with very low and high $\Mstar/\Mhalo$) tend to
preserve the initial cuspy profile, while haloes with
$\Mstar/\Mhalo\approx 0.3\%$ have a flat central density profile.

In this paper we want to expand the original work of DC14, by using a
newer (and larger) sample of simulated galaxies. This new set of
simulations has been performed using the new cosmological parameters
from the Planck satellite (the Planck Collaboration 2014), and an
updated version of the {\sc gasoline} code (Keller \etal 2014) which
fixes some of the known problems of the SPH technique (Agertz \etal
2007).  This newer and larger sample allows us to look at the
evolution of the density profiles through cosmic time and to witness
the creation (and destruction) of density cores as a function of
stellar mass and halo mass.

This paper is organized as follows: In \S2 we give an overview of the
cosmological hydrodynamical simulations used in this work. Results are
presented in \S3, including core creation in \S3.2 and core
destruction in \S3.3. Our conclusions are given in \S4.

\section{Simulations} \label{sec:simulation}

In this study we use simulations from the NIHAO suite (Wang \etal 2015),
based on an updated version of the MaGICC project (Stinson et
al. 2012).  All the simulations have adopted the latest compilation of
cosmological parameters from the Planck satellite (the Planck
Collaboration 2014); namely $\OmegaM=0.3175$, $H_0=67.1$,
$\sigma_8=0.8344$, $n=0.9624$ and $\Omegab=0.0490$.  The haloes to be
re-simulated with baryons and higher resolution have been extracted
from three different pure N-body simulations with a box size of 60, 20 and
15 \Mpch respectively, more information on the collisionless
simulations can be found in Dutton \& Macci\`o (2014).

The aim of the NIHAO project is to study galaxy formation over a large
mass range, from dwarf galaxies to massive spirals such as the  Milky
Way. We have decided to keep the same relative mass resolution
accross the whole mass range, meaning that at high resolution we have
(roughly) the same number of dark matter particles within the virial
radius of our galaxies. This requirement sets the mass of the dark
matter particle and the initial mass of the gas particles, since the
latter is simply obtained by the dark matter mass multiplied by
$(\OmegaM-\Omegab)/\Omegab$.

The zoomed initial conditions were created using a modified version of
{\sc grafic2} (Bertschinger 2001, Penzo \etal 2014). The starting
redshift is  $z_{\rm start}=99$, and each halo is initially simulated
at high resolution with DM-only using {\sc pkdgrav} (Stadel
2001). More details on the sample selection can be found in Wang \etal
(2015).  We refer to  simulations with baryons as the {\it hydro}
simulation, while we will use the term {\it N-body} for the DM-only
simulation.  Table\ref{tab:gal} lists key properties for three
galaxies across the mass range that are analyzed in  more detail; the
complete list of NIHAO galaxies can be found in Wang \etal (2015).

The hydrodynamical simulations are evolved using an improved version
of the SPH code {\sc gasoline} (Wadsley \etal 2004).  The code
includes a subgrid model for turbulent mixing of metals and energy
(Wadsley \etal 2008), heating and cooling include photoelectric
heating of dust grains, ultraviolet (UV) heating and ionization and
cooling due to hydrogen, helium and metals (Shen \etal 2010).

For the NIHAO simulations we have used a revised treatment of
hydrodynamics as described in Keller \etal (2014) that we refer to as
{\sc esf-Gasoline2}.  Most important is the Ritchie \& Thomas
(2001) force expression that improves mixing and shortens the
destruction time for cold blobs (see Agertz \etal 2007).  {\sc
  esf-Gasoline2} also includes the time-step limiter suggested
by Saitoh \& Makino (2009), which is important in the presence of
strong shocks and temperature jumps.  We also increased the number of
neighbor particles used in the calculation of the smoothed
hydrodynamic properties from 32 to 50.

\subsection{Star formation and feedback}

The star formation and feedback modeling follows what was used in the
MaGICC simulations (Stinson \etal 2013).  Gas can form stars  when it
satisfies a temperature and a density threshold: $T< 15000$ K and
$n_{\rm th} > 10.3$ cm$^{-3}$.  Stars can feed energy back into the
ISM via blast-wave supernova (SN) feedback (Stinson \etal 2006, see
Agertz \etal 2013 for a discussion of the implementation of
different  energy and momentum stellar feedback in galaxy formation
simulations)  and
via ionizing radiation from massive stars before they turn in SN.
Metals are produced by type II and type Ia SN.  These, along with
stellar winds  from asymptotic giant branch stars also return mass to
the ISM.  The metals affect  the cooling function (Shen \etal 2010)
and diffuse between gas particles (Wadsley \etal 2008).  The fraction
of stellar mass that results in SN and winds is determined using  the
Chabrier (2003) stellar Initial Mass Function (IMF).  

There are two small changes from the MaGICC simulations.   The change
in number of neighbors and the new combination of softening length and
particle mass means the threshold for star formation increased from
9.3 to 10.3 cm$^{-3}$. The increased hydrodynamic mixing necessitated
a small increase of pre-SN  feedback efficiency, $\epsilon_{\rm ESF}$,
from 0.1 to 0.13.   This energy is ejected as thermal energy into the
surrounding gas, which does not have its cooling disabled.  Most of
this energy is instantaneously radiated away, and the effective
coupling is of the order of 1\%.  

\subsection{NIHAO galaxies stellar masses}

Since our aim is to study the impact of galaxy formation on the dark
matter distribution it is very important to use realistic simulated
galaxies, i.e. galaxies able to reproduce the observed scaling
relations.
Haloes in our simulations were identified using 
halo finder \texttt{AHF}\footnote{http://popia.ft.uam.es/AMIGA}
(Knollmann \& Knebe 2009; Gill \etal 2004), adopting a spherical density contrast
of 200 times the cosmic critical matter density. We dubbed the size of
such a sphere $R_{\rm vir}$, finally  the stellar mass,
$M_{\rm star}$, is measured within a sphere of radius, $r_{\rm gal}\equiv0.2
R_{\rm vir}$.

\begin{figure}
  \includegraphics[width=0.49\textwidth]{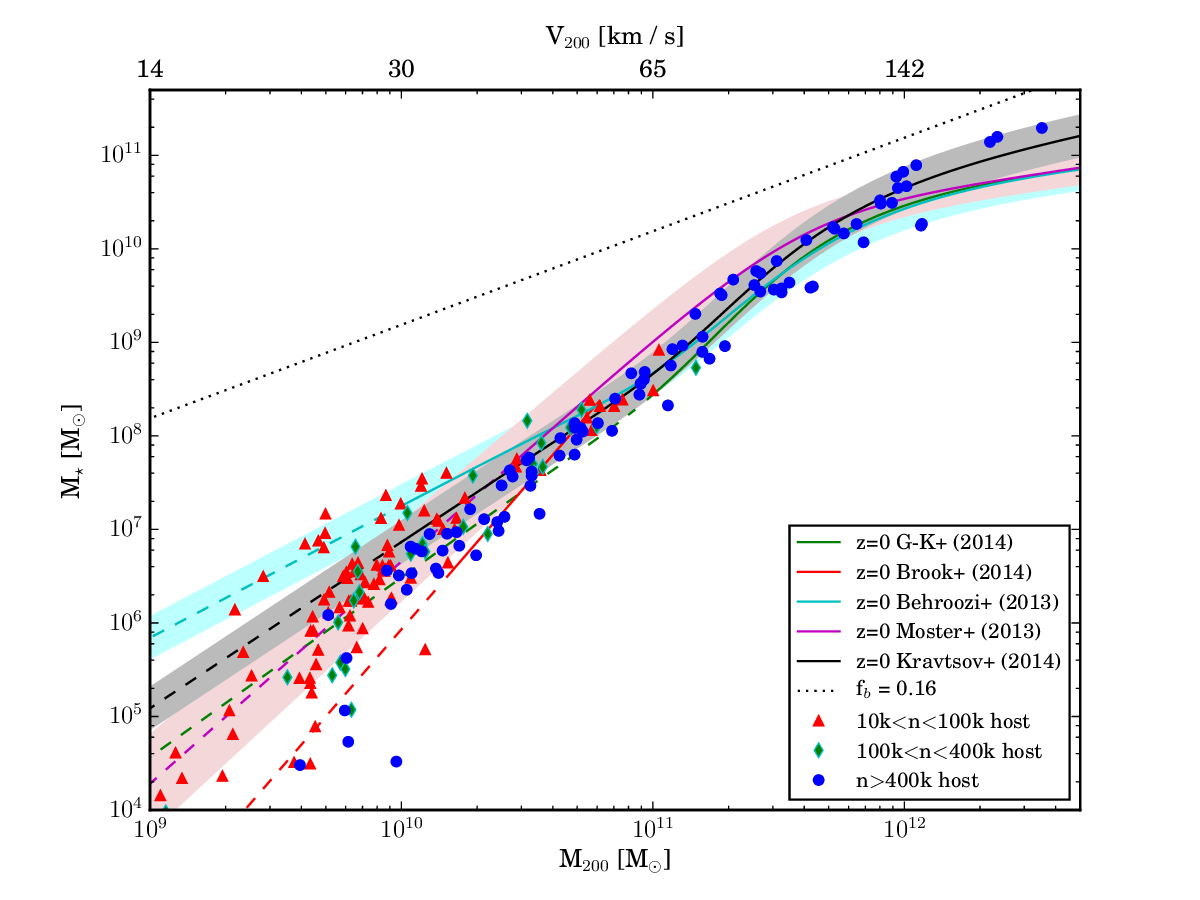}
  \caption{Stellar mass - halo mass relation for the NIHAO galaxies
    used in this work. All simulations have more than 400,000
    particles in their virial radius (see Wang \etal 2015 for more
    details). The solid lines represent the most commonly used
    abundance matching results (see text).  }
\label{fig:am}
\end{figure}

Fig.~\ref{fig:am} shows the stellar mass - halo mass
relation for the NIHAO galaxies used in this paper, compared with the
most commonly used abundance matching relations from the literature:
Brook \etal (2014), Kravtsov \etal (2014), Garrison-Kimmel \etal (2014), Behroozi \etal
(2013) and Moster \etal (2013).  This plot is similar to the one
showed in Wang \etal (2015) with the only difference that here only
the most resolved halo in each simulation is shown. As detailed in
Wang \etal NIHAO galaxies are able to reproduce the the stellar mass -
halo mass relation also at higher redshift, and have realistic star
formation rates for their stellar masses.  Overall the unprecedented
combination of high resolution and large statistical sample of the
NIHAO suit offers a unique tool to study the response of DM to galaxy
formation.

\subsection{Profile fitting}

To construct and fit the dark matter density profiles we used the same
methodology as introduced in DC14.  Each Dark Matter density profile
(in both hydro and N-body simulations) has been computed using
regularly spaced shells on a logarithmic scale.  The center of the
halo has been determined using the shrinking sphere method (Power et
al. 2003) and we set the minimum radius for the first shell as twice
the softening length of the dark matter particles.  Each density
profile is computed up to the virial radius, which is determined using
the spherical overdensity criterion with a virial overdensity of
$\Deltavir=200$ times the critical density of the universe.  The error
on the density in each bin is set according to the Poisson noise due
to the finite number of particles each density bin.

Following Di Cintio \etal (2014a) the DM  central  density  slope
($\alpha$) is  subsequently fitted  using  a  single power law, $\rho
\propto r^{\alpha}$, over a limited radial range, namely
$0.01<r/R_{\rm vir}<0.02$, where $R_{\rm vir}$ is the virial radius.

\subsection{Profile convergence}

The determination of the minimum radius above which the results
of a simulation are not affected by the finite resolution is a non
trivial task (e.g. Gao \etal 2008).
A quite commonnly used criterion for convergence has been suggested
by Power \etal (2003) for collisionles simulations, and it is based
on the two body relaxation time scale for particles
in a gravitational potential.
This empirical criterion radius ensures that the mean density inside the convergence
radius is within 10 \% of the value obtained in a simulation of much
higher resolution (Schaller \etal 2015).

Due to choice of having similar particle resolution in all the NIHAO galaxies, this convergence radius
is quite constant across the whole mass range and at $z=0$ is $\approx0.4-0.7$\%  of $R_{\rm vir}$.
At $z=2$, the inner region is marginally resolved if we use the original definition
of Power \etal 2003, since we obtain a convergence radius of about 2\% of the virial radius.
On the other hand as noticed by Shaller et al. (2015), this definition is quite restrictive
and it was based on pure DM simulations. For hydrodynamical simulations it can be relaxed
by requiring a 20\% convergence in the enclosed density, (instead of 10\%).
If we follow the prescription of Schaller \etal the convergence radius is of the order
1.0-1.5\% of $R_{\rm vir}$, this means that while being ``on the edge'' we can still use the same
defintion for $\alpha$ up to $z=2$, as long as we do not try to be too quantitative at
these redshifts.

\section{Results}\label{sec:results}

Thanks to the large range of masses covered by our simulations we are
able to see both expansion and contraction at $z=0$ of the dark
matter profile, depending on the halo mass and stellar content (see
DC14 and references therein).

\begin{figure}
  \includegraphics[width=0.49\textwidth]{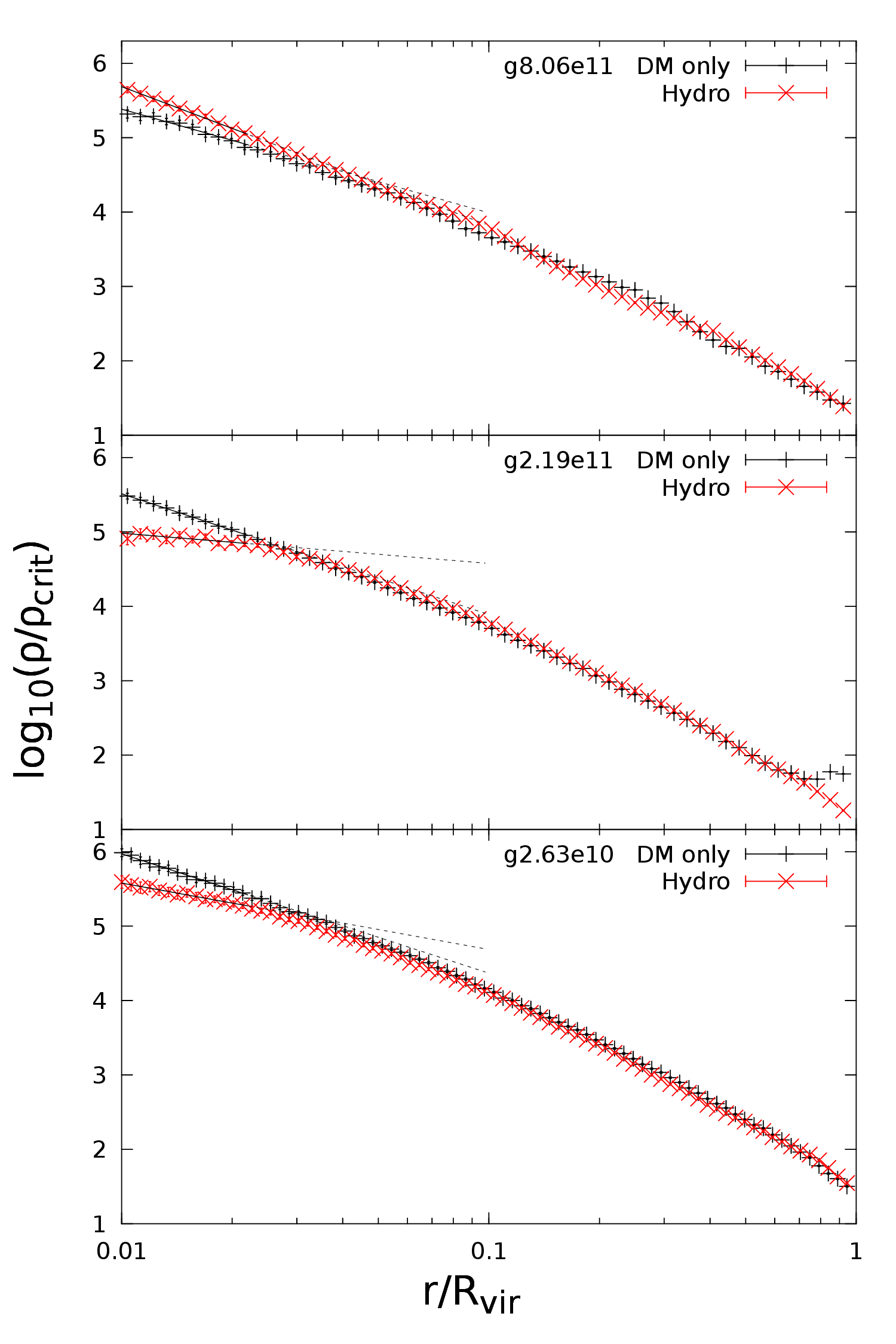}
  \caption{Density profiles for three different simulations; on each
    plot we present the profile at $z=0$ for the collisionless N-body
    (in black) and hydrodynamic (in red) simulation. The lines show
    power-law fits to the inner 1-2\% of the halo, and are shown to
    0.1$\Rvir$ for clarity. The upper panel
    shows an example where the DM halo contracts (halo mass of
    $9.43\times 10^{11}\Msun$ and stellar mass of $4.48\times
    10^{10}\Msun$), the middle panel shows a galaxy with a DM
    halo that expands (host halo mass of ),  
    while the bottom panel shows a galaxy with very little
    change (halo mass of $2.70\times 10^{10}\Msun$ and stellar mass of $4.28\times
    10^{7}\Msun$).}
\label{fig:DMz=0}
\end{figure}

\begin{figure*}
  \includegraphics[width=0.95\textwidth]{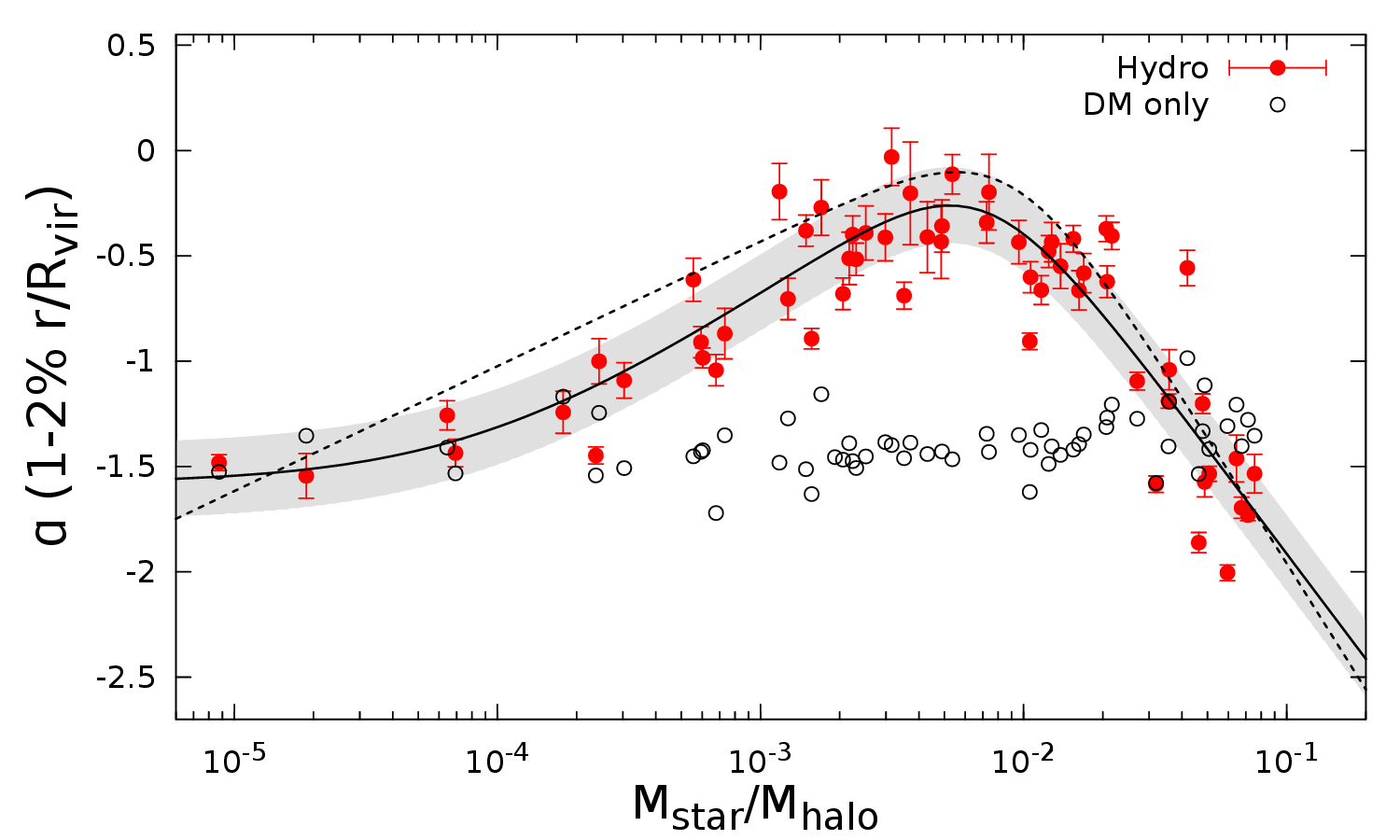}
  \caption{Inner dark matter density slope, $\alpha$, as a function of
    $\Mstar/\Mhalo$ for galaxies at $z=0$. The filled red circles show
    results for hydro runs (with errorbars from the profile fit) while the black open circles show results
    for the N-body DM only runs. The dotted line is the original
    function from Di Cintio \etal (2014a), the thick line is the
    refitted curve with our new simulations using the Planck cosmology
    with the shaded region showing the 1$\sigma$ scatter of 0.18.}
\label{fig:alphaz=0}
\end{figure*}

An example of such a dichotomy is presented in Fig. \ref{fig:DMz=0},
where we show a comparison of the dark matter density profile at $z=0$
in the hydro and in the N-body run for three galaxies with different
total masses.  
The middle panel of Fig.~\ref{fig:DMz=0} shows the DM density profile
for a galaxy with a virial mass of $\approx 2.2\times 10^{11} \Msun$.
The profile in the hydro case (red points) shows a quite
extended core.  The situation is reversed for a more massive halo
($\approx 9.4\times 10^{11} \Msun$, upper panel) where the DM profile
is contracted in the hydro simulation with respect to the N-body run (black points).
The lower panel shows a low mass galaxy (halo mass $\approx 2.7\times 10^{10}
\Msun$), in this case the both profiles are cuspy, but it is clear
that the hydro simulation has a shallower central slope than
the N-body one.

This plot shows that simulations  that include a realistic treatment of
hydrodynamics do not share a  universal dark matter density profile in
contrast with dark  matter only simulations.  We go  into detail about
how the universality is broken in section 4.

As already pointed out by Di Cintio \etal (2014a) there is tight
relation between the DM response and the star formation efficiency of
a galaxy, defined as the ratio between stellar mass and halo mass.
Fig.~\ref{fig:alphaz=0} shows the relation between $\alpha$ and the
$\Mstar / \Mhalo$ ratio. The red points with errorbars are the NIHAO
simulations results at $z=0$. The black open circles are the results
from the N-body runs, and for these simulations we use the stellar
mass of their hydrodynamical counterpart.

The NIHAO results show a quite interesting behaviour: the inner slope,
$\alpha$, varies as a function of star formation efficiency.  At low
values of $\Mstar / \Mhalo$ the N-body and hydro results are quite
similar and both predict a constant value of $\alpha \sim -1.5$.  Then
$\alpha$ increases steadily and reaches a maximum for  $\Mstar /
\Mhalo \sim 6 \times 10^{-3}$, in agreement with previous findings
from DC14 (black dashed line).  At the highest masses, after the
maximum expansion, $\alpha$ decreases and for very large values of
$\Mstar / \Mhalo$ of the order of few percent, the hydro slopes become
steeper than their N-body counterparts, suggesting halo contraction on
those scales. This is in agreement with previous results from Di
Cintio \etal 2014b, where it was also found that at the mass scales of
the Milky Way, the DM profiles have a much higher concentration
parameter in hydro simulations compared to collisionless results.  It
is also worth remembering that, as shown in Fig.~\ref{fig:am}, all our
galaxies are on the observed stellar mass-halo mass relation
(e.g. Behroozi \etal 2013; Kravtsov \etal 2014), and hence our halo
contraction is not due to unphysical overcooling as it happened in
older simulations (e.g. Gnedin \etal 2004) and it is consistent with
recent results from other groups on the same mass scale (Schaller
\etal 2015).

We tried to capture the behaviour of the slope of the density profile
as a function of the star formation efficiency ($x=\Mstar / \Mhalo$)
with a different fitting formula than DC14:
\beq
\alpha(x)= n - {\rm log}_{10}\left[n_1 \left(1+\frac{x}{x_1}
\right)^{-\beta}+\left(\frac{x}{x_0}\right)^\gamma\right],
\label{eq:fit_function}
\eeq

This new fitting function has the advantage that it converges to a
fixed value ($\approx -1.5$) for $x$ that goes to zero, in contrast to
the power-law behavior of the initial suggestion from DC14.  Since,
for continuity, we do expect to recover the slope of the N-body results
for a negligible amount of stars ($x\lta 10^{-5}$),
this new formula provides a better fit to the simulation data. The
grey area is the one sigma scatter around the mean
value of $\alpha$ of $\approx 0.18$.

\begin{table}
\label{tab:fit}
\caption{Best fit parameters for the value of $\alpha$ computed
  witihin 1 and 2 \% of \Rvir as a function of $\Mstar/\Mhalo$, $\Mhalo$
  and $\Mstar$.}
\begin{center}
\begin{tabular}{lcccccc}
\hline
\hline
 & n & $n_1$ & $x_0$ & $x_1$ &  $\beta$ &$\gamma$ \\
\hline
${M_*} \over {M_h}$ & -0.158 & 26.49 & 8.77 $\times 10^{-3}$ & 9.44$\times 10^{-5}$ & 0.85 & 1.66 \\
   \hline
${M_h} $ & 0.70 & 99.14 & 9.94$\times 10^9$ & 9.80$\times 10^{10}$ & 4.85   & 1.00 \\
  
   \hline
${M_*}$ &  0.53 & 61.44 & 2.48$\times 10^6$ & 6.83$\times 10^8$ & 29.5 & 0.42 \\
   \hline
\end{tabular}
\end{center}
\end{table}

The difference between the two curves for $\Mstar / \Mhalo > 10^{-3}$
is mainly due to the different cosmological model adopted in the two
studies (Planck vs. WMAP3), while other differences, as for example the
improved hydrodynamics, are just a secondary effects
(see Stinson \etal 2015 for a more detailed comparison between
MaGICC and NIHAO galaxies).
In the Planck cosmology the halo
concentrations are higher by $\approx 45\%$ (Dutton \& Macci\`o 2014),
which explains why we get lower (cuspier) $\alpha$-values with respect to DC14.
Table \ref{tab:fit} lists the values of our new relation for $\alpha$
vs.  $\Mstar / \Mhalo$, (updated to the Planck Cosmology).

\begin{figure}
  \includegraphics[width=0.49\textwidth,angle=0]{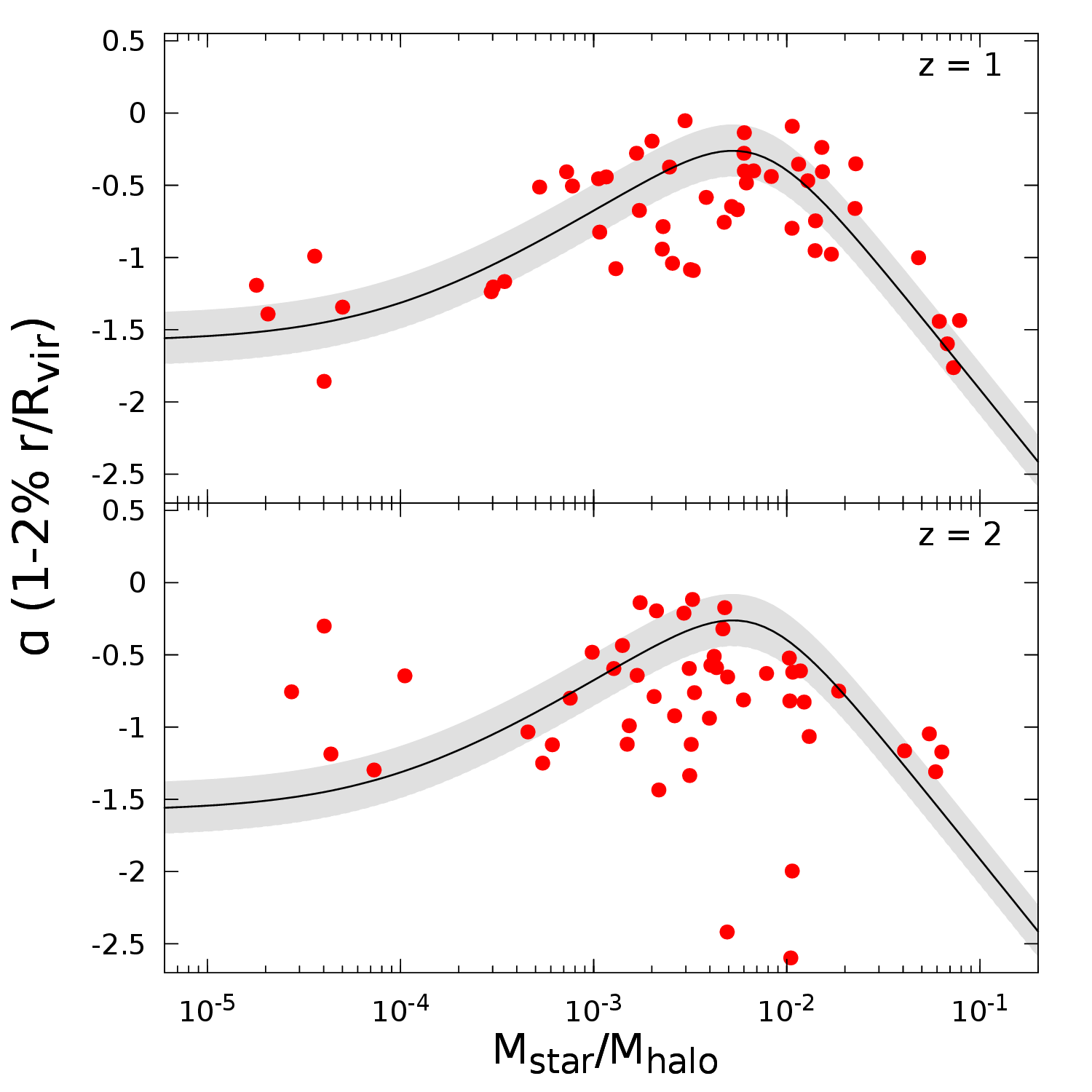}
  \caption{Inner dark matter density slope, $\alpha$, as a function of
    $\Mstar / \Mhalo$ for galaxies at $z = 1$ (upper panel) and at
    $z=2$ (lower panel). The solid line and shaded region is $z=0$ relation from
    Fig.~\ref{fig:alphaz=0}.}
\label{fig:alphaz=2}
\end{figure}

In Fig.~\ref{fig:alphaz=2} we show the same relation as in
Fig.~\ref{fig:alphaz=0} but at redshift $z=1$ (upper panel) and at
redshift $z=2$ (lower panel). The line in the plot is still the $z=0$
fitting result. 
As the galaxy evolve, dark matter accretion and star formation generally
move  points from  left to  right.  The points still cluster around the $z=0$ 
relation, but scatter  increases with redshift, so  
integrated star formation efficiency  becomes a less  
useful metric when the galaxies are young.

The origin of the scatter is two fold. Part of it is due to resolution
effects, at higher redshift the virial radius is smaller, this means
that we are probing regions closer to our resolution limit which
implies a lower number of particles in the inner region of the halo
and hence a larger Poisson noise.  Part of the scatter is also
physical and related to the bursty nature of the star formation
history at high redshift, that induces short time variations in the
density profile (see next section). 

\begin{figure}
  \includegraphics[width=0.49\textwidth,angle=0]{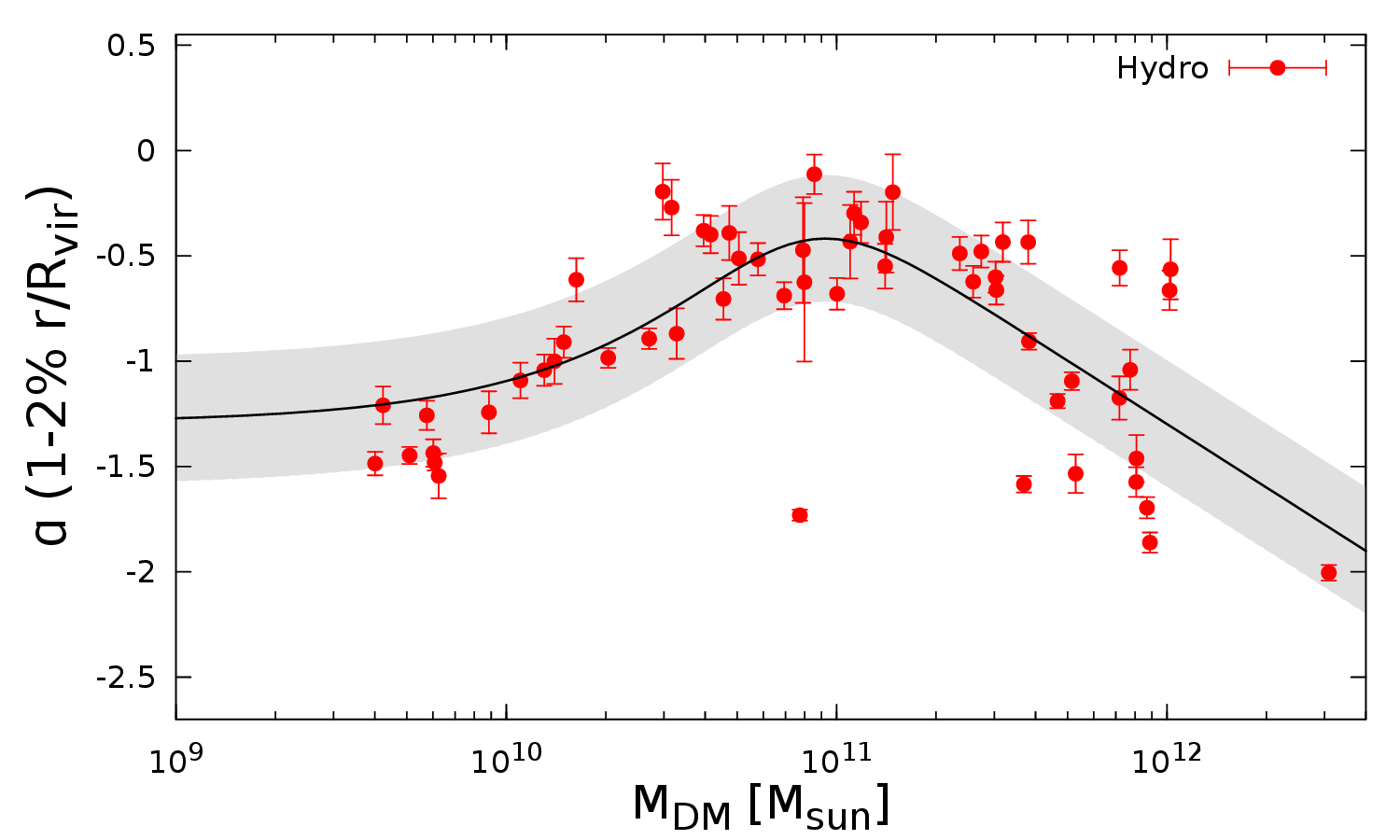}
  \caption{Inner dark matter density slope, $\alpha$, as a function of
    the halo mass for galaxies at $z = 0$. The solid line and shaded 
    region are the fitted relation using Eq.~\ref{eq:fit_function} and 
    its scatter. The parameters of the fitting are reported
    in Table \ref{tab:fit}.}
\label{fig:alphaMDM}
\end{figure}

\begin{figure}
  \includegraphics[width=0.49\textwidth,angle=0]{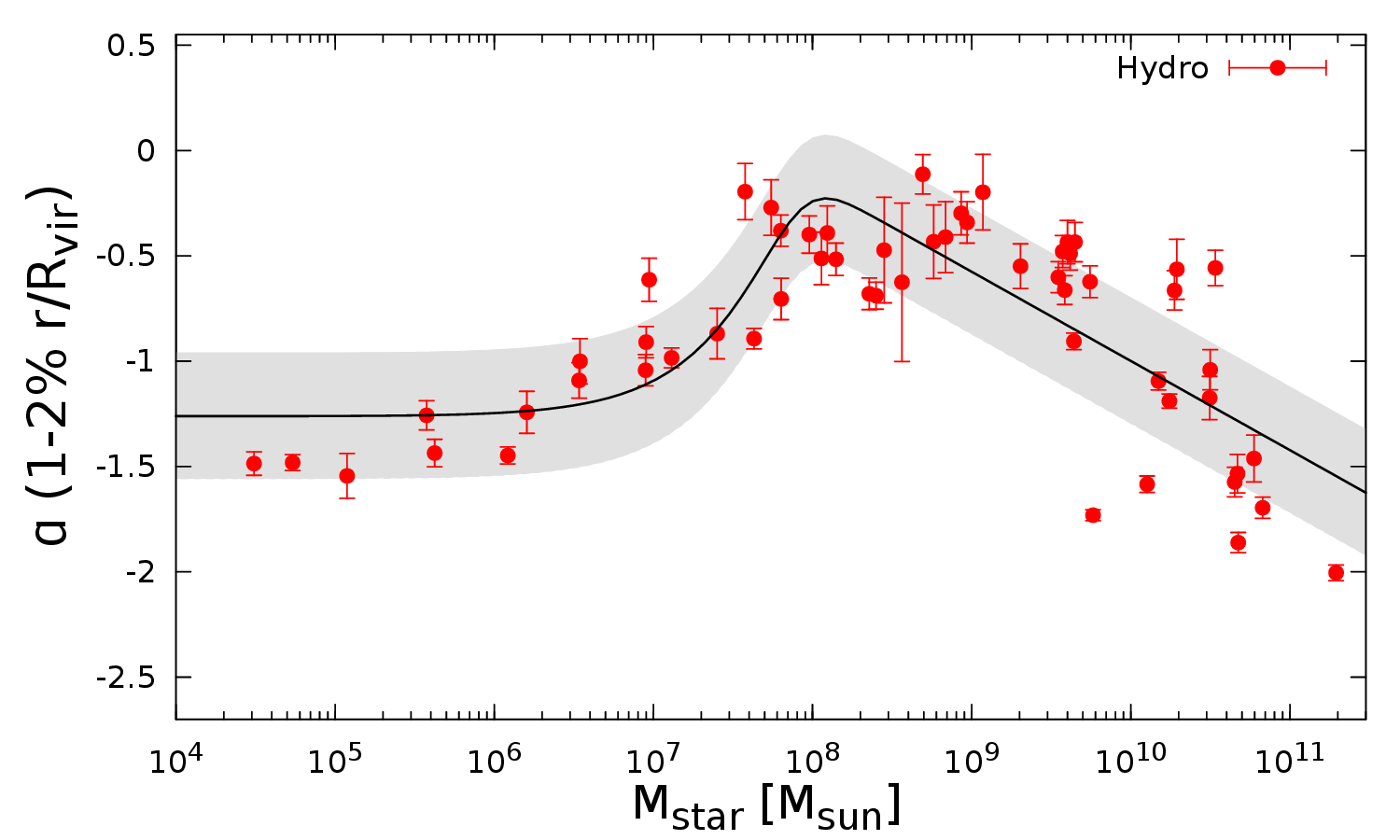}
  \caption{Inner dark matter density slope, $\alpha$, as a function of
    the stellar mass for galaxies at $z = 0$. The solid line and
    shaded  region are the fitted relation using
    Eq.~\ref{eq:fit_function} and its scatter. The parameters of the
    fitting are reported in Table \ref{tab:fit}.}
\label{fig:alphaMS}
\end{figure}

To check whether star  formation efficiency  best  correlates with
$\alpha$  in  our large  sample  of galaxies, Fig. \ref{fig:alphaMDM}
and  Fig. \ref{fig:alphaMS} show how  $\alpha$ varies with the halo
mass and the stellar mass  of our galaxies respectively.  In both
cases the behaviour of $\alpha$ is similar to the one seen in
Fig.~\ref{fig:alphaz=0}, with a maximum in 'core' creation around
$10^{11}$ $(10^8)$ \Msun in halo (stellar) mass.  These plots 
show a similar scatter in  $\alpha$ as in Fig.~\ref{fig:alphaz=0}
for low values of the halo (stellar) mass: $M_{\rm halo}<5\times 10^{11} \Msun$
($M_{\rm star} <10^9 \Msun$), but a larger scatter for $\alpha$ at high masses.
This confirms the earlier results of Di
Cintio \etal (2014a) where using galaxies run with different stellar
feedback prescription, they show that the integrated star formation
efficiency is the best parameter to capture the effect of baryons on
the DM distrubution.

The role of the integrated star formation efficiency is also important
to explain the similiraties and (small, partial) differences between our results
and some previous studies on the same topic from different groups.

Madau \etal (2014) analyzed the
evolution of the density profile of a small group of dwarf galaxies,
with only 4 of them with containing stars.
The simulations were run with an earlier version of the {\sc gasoline} code,
with similar SN feedback.
Two of those galaxies
(dubbed Doc and Bashful) have masses around $1-3 \times 10^{10} \Msun$ and
do show extended cores, somehow in contrast with the results shown in figure
\ref{fig:alphaMDM}. On the other hand the galaxies presented by Madau and
collaborators have over-massive stellar bodies, that over predict results
from Local Group abundance matching from  Brook \etal (2014) and Garrison-Kimmel \etal (2014)
by a factor  $>10$ for ``Doc'' and of about
$3-4$ for ``Bashful'' (w.r.t Garrison-Kimmel \etal even more for Brook \etal).
This overcooling can be explained by the lack
in their simulation of any other source of feedback besides  SuperNovae
(see for example the results of Stinson \etal 2013).
When the higher (may be too high) stellar mass -halo mass ratio
of these galaxies,  compared to our NIHAO sample, is taken into account
the results are in very good in agreement.

A similar agreement is also found when comparing our rusults
with recent simulations presented by the FIRE collaboration in Chan \etal (2015).
For example there is a quite remarkable similarity in the peak of the core
formation and in the non monotonic response of DM to galaxy formation.
This similarity is even more striking when we take into account
that the two simulation sets (NIHAO and FIRE) differ in resolution,
numerical technique and stellar feedback implementation.

One place where our and their results do not agree is in the decline
of the value of $\alpha$ at low masses and which
brought them to suggest  the possibility of a ``threshold'' halo mass at
around $10^{10} \Msun$ needed to develop large cores.
It is worth mentioning that while the FIRE simulations reach a much higher resolution
than our current sample, they don't have large number statistics,
with only  9 galaxies across 5 orders of magnitude in stellar mass.

We do not see such a threshold in our sample at this halo mass (but note
that we do see a substantial change in $\alpha$ for a stellar mass of $\approx 10^7$ \Msun).
This difference can have many explanations.
The higher resolution of the FIRE simulations might model the interplay between star formation and the outflows they drive better
The partial evidence for a threshold could be due to the much lower sample size of Chan \etal: they only
have three galaxies in the dark matter range between $5\times 10^9$ and $5\times 10^{10} \Msun$,
while we have twenty five, and hence we are less subject to artificial ``jumps'' due to low number statistics
and intrinsic scatter.

Finally as in the case of the Madau \etal results, the FIRE simulations predict
a stronger star formation efficiency at low masses, which partially over produces
abundance matching results from Garrison-Kimmel \etal (2014) and Brook \etal (2014).
This departure from abundance matching results, starts exactly above halo masses of $10^{10}$ \Msun
(Dutton \etal 2015, figure 1). Overall we would like to stress again the importance
of having large samples and realistic stellar to halo masses ratios in order to infer the behavior
of the dark matter distribution in hydrodynamic simulations.

\subsection{Single profile evolution}

\begin{figure}
  \includegraphics[width=0.49\textwidth]{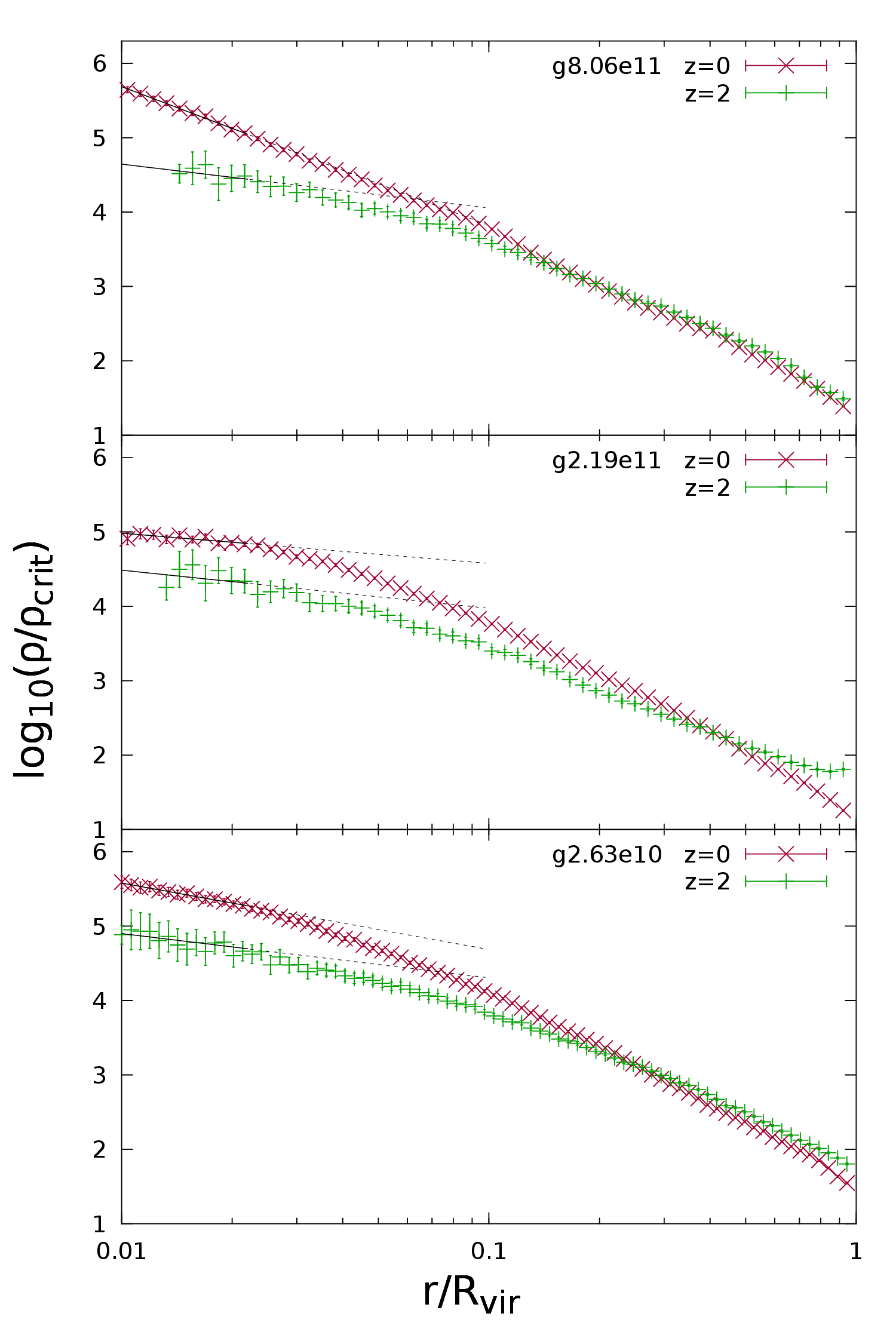}
  \caption{DM density profiles for our three test  galaxies at two
    redshifts. The profiles at $z = 0$ are
    represented in red, and at $z = 2$ in green. The dotted line is
    fitted for each profile between one and two percent of the virial
    radius and is shown up to ten percent of the virial radius to help
    visualize.}
\label{fig:profz}
\end{figure}

To study  the evolution of  individual halos,  we make
case studies  of 3 simulations,  g2.63e10, g2.19e11, and g8.06e11 (already
shown in  Fig. \ref{fig:DMz=0} and presented in table \ref{tab:gal}), that  show 
different behaviors as a function of redshift.
In Fig. \ref{fig:profz}, we show the DM
density profiles at $z=0$ (red points) and $z=2$ (green points) for the
three galaxies. At $z=2$ the total halo masses are $1.48 \times 10^{10}$
$2.61 \times 10^{10}$ and $1.50 \times 10^{11}$, 
while the stellar masses are $3.29 \times 10^{7}$, $2.64 \times 10^{7}$,
and $1.11 \times 10^{9}$, for g2.63e10 , g2.19e11, and  g8.06e11 respectively.
The dashed line shows a power-law fit to the density between 1 and 2\% 
of the virial radius.

As expected, at $z=2$  the middle mass galaxy already has an expanded
profile (compared to the N-body run), with a slope of $\alpha=-0.65$,
and then the inner density flattens  by the end of the simulation.  At
$z=2$, the dark matter profile of the higher mass halo looks nearly
identical to the  middle  mass halo at  $z=0$. It evolves differently;
its central density profile becomes steeper with time.
The low mass galaxy (g2.63e10) has a very mild evolution and the slope
of the profiles doesn't change much from $z=2$ to the present day.

This strong time variation of the size of the central density core are
correlated to quick variations in the Star Formation Rate of the
galaxy (Pontzen \& Governato 2012, see also Macci\`o \etal (2012) for the effect of the sloshing of the gas inside
the potential well).  

\begin{figure}
  \includegraphics[width=0.48\textwidth]{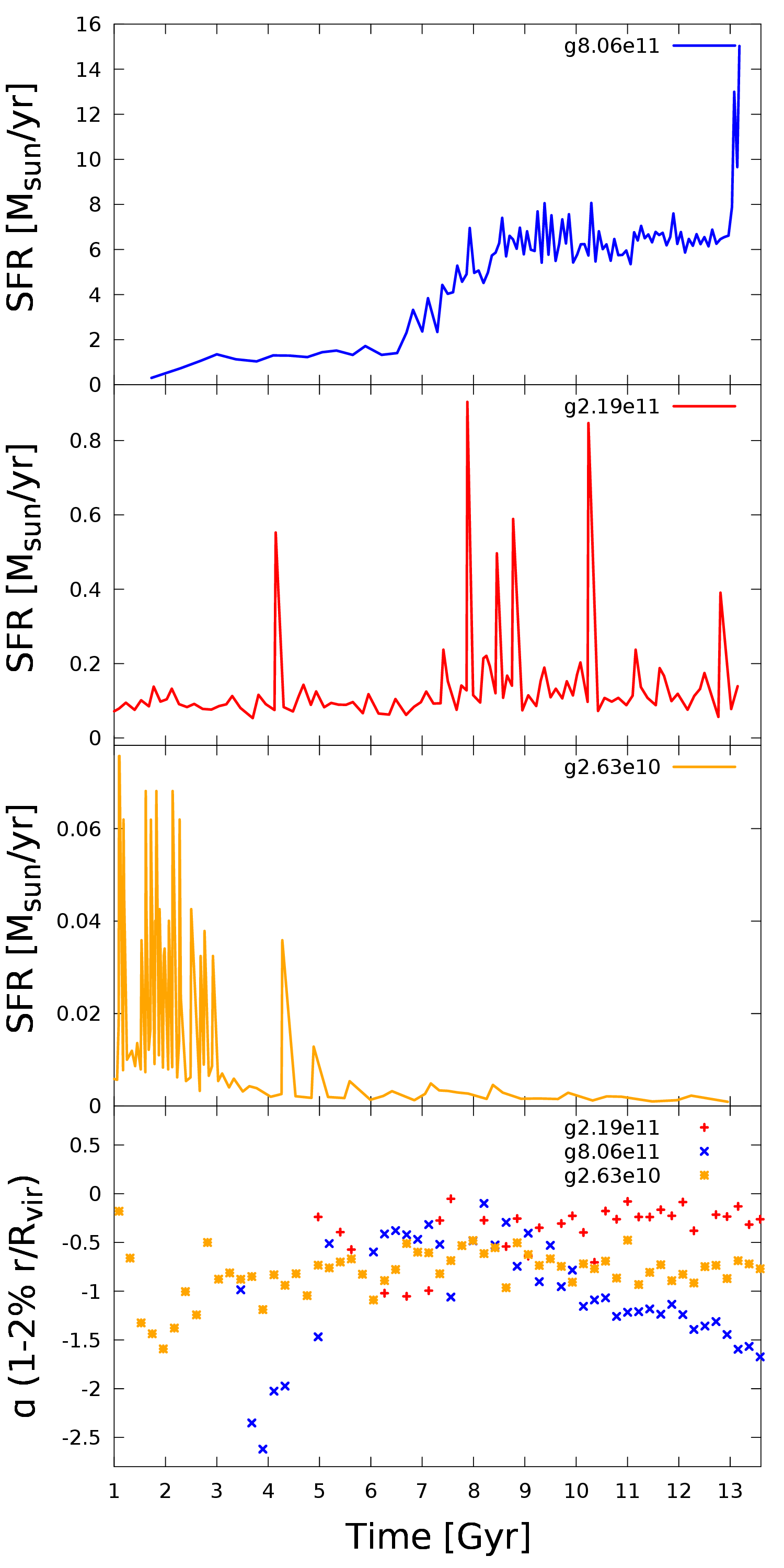}
  \caption{Time evolution of the Star Formation Rate (SFR, upper
    panels) and the central density slope ($\alpha$, lower panel), for
    our three test galaxies: g8.06e11 (blue), g2.19e11 (red), g2.63e10
    (orange).} 
\label{fig:4pan}
\end{figure}

Fig. \ref{fig:4pan}  provides example  cases of  how bursty  star
formation  effects the  inner dark matter profile slope  for galaxies
of three  different masses spanning the NIHAO sample.  The lowest mass
case, g2.63e10, has early bursts of star formation  before its star
formation dies  away entirely.  The bursts of star formation are not
reflected in the inner profile slope at  early times,  most likely
because of  the low  resolution of  the profiles that makes it hard to
measure the slope.  As time goes on and the resolution  increases,
the slope  stays moderately  flatter than NFW, at a value slightly
higher than $-1$.

The middle mass  case, g2.19e11, displays bursts  of star formation
throughout  its evolution.   The  largest bursts  start  around 7  Gyr
($z\sim1$).  The bursts  coincide with a flattening of  the inner dark
matter profile slope.

The  highest mass  galaxy, g8.06e11,  shows a  monotonically rising
star  formation history.   Throughout  the history,  there are  bursty
periods.  Its  inner density  profile evolves  from a  cusp to  a core
before contracting back  into a cusp with $\alpha=-1.7$  at $z=0$.  It
changes from a cusp to a core at the same time that its star formation
rate increases by a factor of 4, around 7 Gyr.

\subsection{Core creation}

The variation of the central slope of the density profile is also
linked to the creation (and destruction) of a constant density inner
core.  In order to have an estimate of the size of such a core, we
have decided to follow the approach of Macci\`o \etal (2012) and fit
the profiles with the following parametric description, originally
presented in Stadel \etal (2009):
\begin{equation}
\rho (r) = \rho_{0} \exp(\lambda [\ln(1+r/ R_{\rm \lambda})]^2).
\end{equation}
In this parameterization, the density profile is linear down to a
scale $R_{\rm \lambda}$ beyond which it approaches the central maximum
density $\rho_0$ as $r \rightarrow 0$.  We also note that if one makes
a plot of $\rm {d ln} \rho/ {\rm d ln (1 + r/R_{\lambda})}$ versus $
\rm{ln} (1 + r/R_{\lambda})$ then this profile forms an exact straight
line with slope $2\lambda$.  This fitting function is extremely
flexible and makes it possible to reproduce at the same time both
cuspy and cored density profiles. In the following we will use
the fitted value of $R_{\rm \lambda}$ as the core size. 

\begin{figure}
  \includegraphics[width=0.48\textwidth]{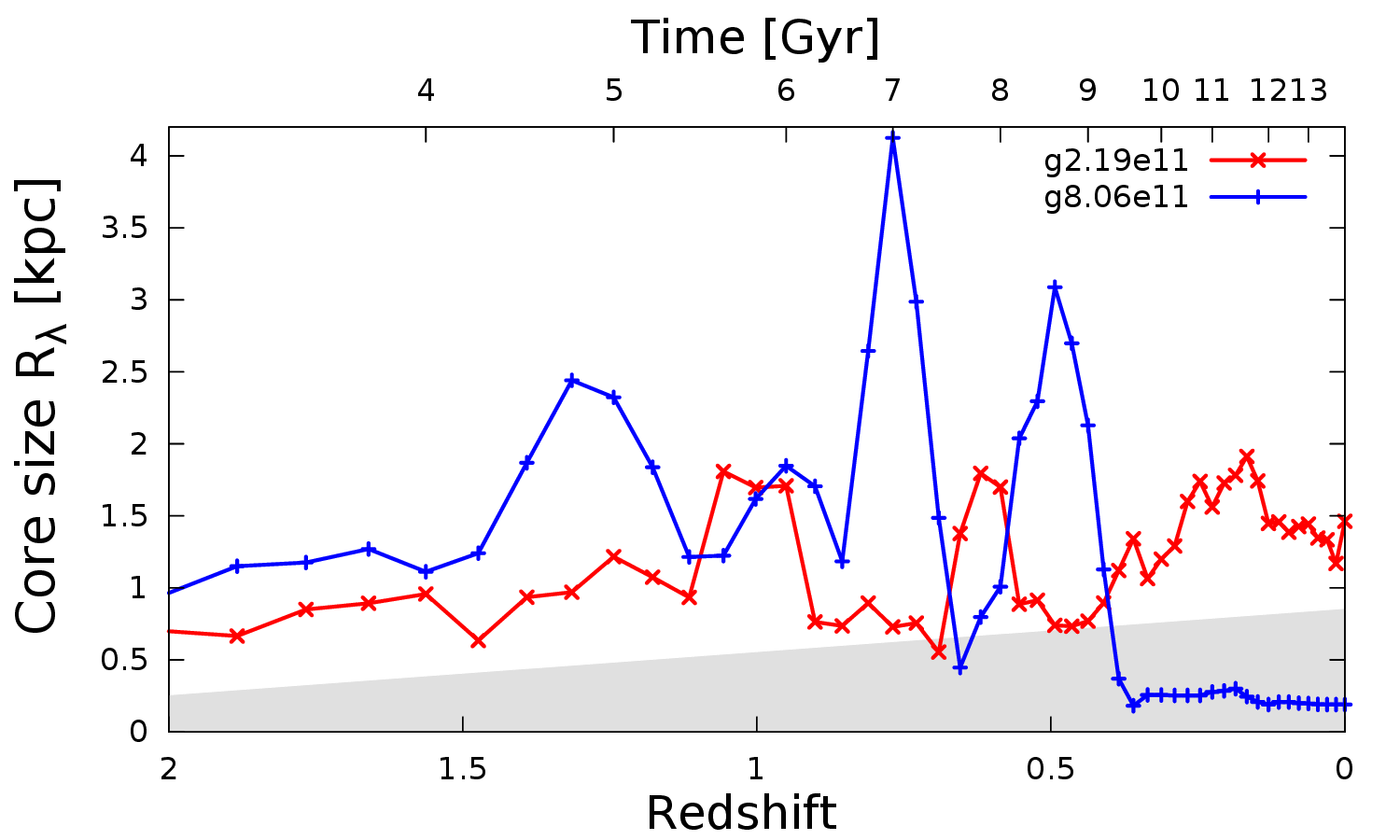}
  \caption{Core size $R_{\lambda}$ as a function of time $1/(1+z)$ for
    g2.19e11 (red) and g8.06e11 (blue). The grey shaded region
    corresponds to the resolution limit of the simulations.}
\label{fig:core}
\end{figure}

In Fig. \ref{fig:core} we show the core size ($R_{\rm \lambda}$) as
function of time for the two more massive galaxies shown in
Fig.~\ref{fig:profz}.  The grey band in the lower panel of the plot
represents the softening of the simulations, any core value below this
limit has to be interpreted as a non-cored profile.
The lower mass galaxy (red line) shows a gradually increasing average
core size from $z=2$ to $z=0$.  The behavior of the more massive
galaxy (blue line) is much more interesting.  At high redshift ($z=2$)
the halo has a core of order $\sim 1.5$ kpc, in between $z=1.5$ and
$z=1$ the core size fluctuates a lot and reaches quite large value up
to 4 kpc. The core is then definitely destroyed after $z=0.5$ and the
galaxy shows no presence of a central constant density, as confirmed
by the profile shown in Fig.~\ref{fig:DMz=0}.

\begin{figure}
  \includegraphics[width=0.49\textwidth]{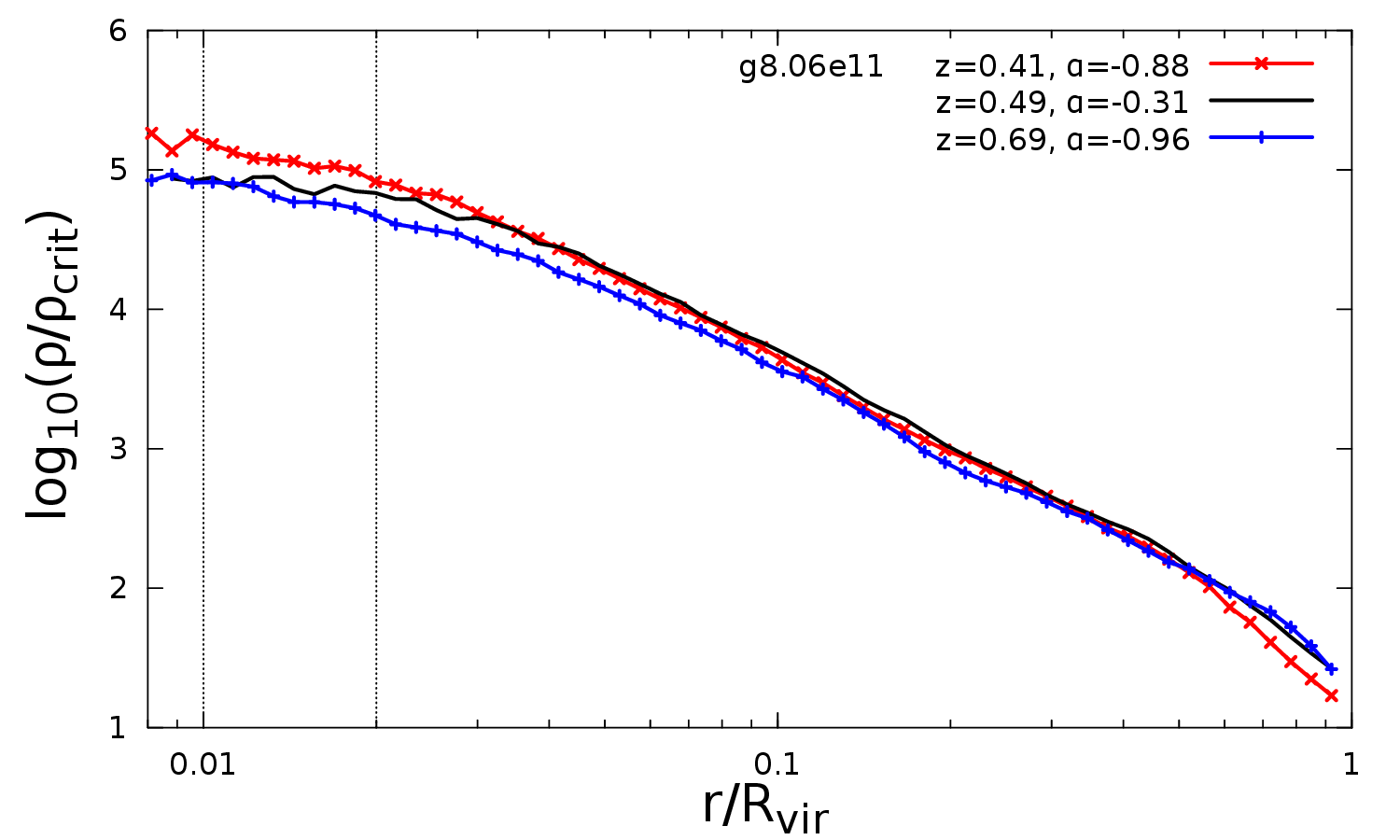}
  \caption{Evolution of the Dark Matter density profile for the
    g8.06e11 halo around $z\sim 0.7$. The profile changes 
    significantly with clear evidence for the creation and
    subsequent destruction of a large density core. The time spanned
    by the three profiles is of the order of one Gyr. The two dotted lines
  mark the region within which $\alpha$ is computed.}
\label{fig:spike}
\end{figure}

It is interesting to look deeper into the origin of the spikes of the
core value around $z\sim 0.5$ ($T\sim8.5$ Gyrs) for the g8.06e11 halo.
In Fig.~\ref{fig:spike} we show the dark matter density profile of this galaxy
at the time of the maximum size of the core ($z=0.67$, black line),
and right before the core creation (red line) and right
after the core destruction (blue line).
It is clear from the figure that the profile does change quite
strongly in the time interval spanned by the spike. A large core (up
to 4 kpc) is created and then destroyed in less than a Gyr.

This fluctuation are related to the rapid increas of an order of magnitude
of star formation from T=6.5 Gyrs to T=8.5.
The morphological analysis of the galaxy shows that this extended burst
is due to  disc instability, possibly triggered
by a satellite accretion.
During this phase a lot of gas is both accreated and ejected
from the central region creating the perfect conditions for
core creation and destruction (e.g. a la Pontzen
\& Governato 2012).

\subsection{Core destruction}

Our simulations clearly show that dark matter halo profiles are not a
static entity. On the contrary they change significantly during the
formation and evolution of the halo. While all haloes start as being
cuspy, as predicted by pure CDM simulations, they can develop a core. 
Depending on the stellar to halo mass
ratio, the core can persist to the present day, or be destroyed, being
replaced by a central cusp.

\begin{figure}
  \includegraphics[width=0.49\textwidth]{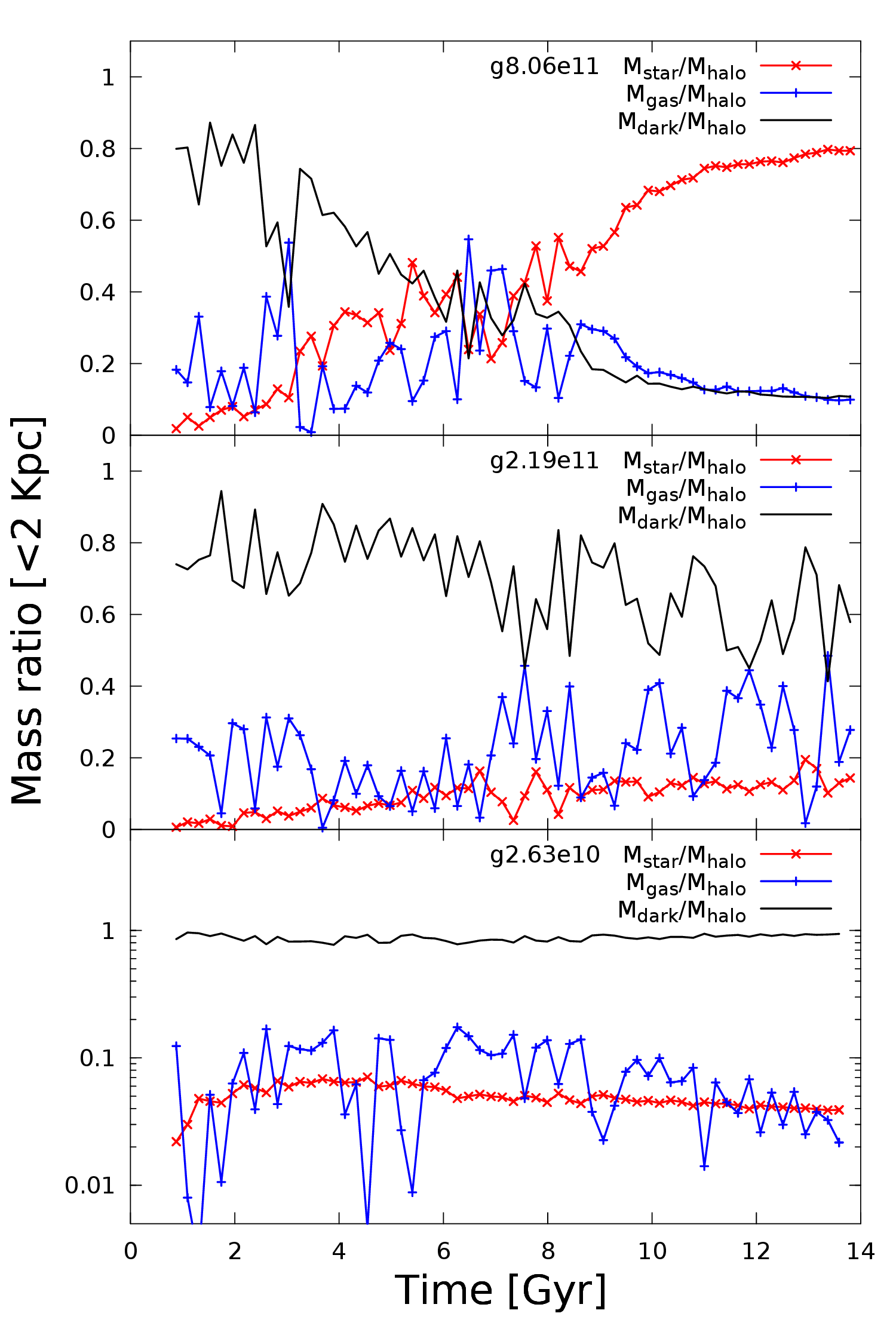}
    \caption{The contribution of DM (black), gas (blue) and stars
      (red) to the total mass in the inner 2 kpc for g8.06e11 (upper
      panel) g2.19e11 (middle panel) and g2.63e10 (lower panel), as a
      function of time. Note the log y-axis in the lower panel. }
\label{fig:central2kpc}
\end{figure}

In  massive haloes (e.g. with halo mass around $10^{12} \Msun$) the
cusp regeneration is mainly due to strong gas inflow in the central
region, which builds up a large stellar body in the center of the
galaxy causing a deepening of the local potential well.  We will now
analyze these processes in more detail.

In Fig.~\ref{fig:central2kpc} we show the relative abundance of dark
matter, gas and stars in the inner 2 kpc for our test galaxies.   In
the case of the lowest mass galaxy (g2.63e10) the central potential is
always  dominated by dark matter which accounts for more than $\sim
90\%$ of the total mass in the center.  The lack of evolution in
$\alpha$ shown in Fig.~\ref{fig:4pan}  is then due to the very
``passive'' evolution of the central region of this halo after star
formation is over. Nevertheless the final effect of baryons on this
low mass halo is still  a {\it net expansion} (as
shown in figure \ref{fig:DMz=0}) with $\alpha=-0.9$ in the Hydro run
with respect to a value of $\sim -1.6$ in the N-body one; despite stars and gas
account for only  $\sim 10\%$ of the mass inside 2 kpc.

For the middle mass galaxy (which retains a cored profile all the way
to $z=0$), the potential in the central region is dominated by dark
matter and gas, while stars are sub-dominant.  The amount of gas is
rapidly varying with time and this causes quick potential variations
and a rapid movement of the center of mass of the gas compared to that
of the dark matter; both these effects contribute in creating and
maintaining a density core.

The situation is quite different for the most massive galaxy: after
$T=8$ Gyr the potential in central region is dominated by stars with
the gas (and DM) being highly sub-dominant. At this point, the deeper
potential well caused by the stars is anchoring the global potential
and making any possible gas outflow a negligible effect.  The dark
matter reacts to the new, enhanced central potential by contracting
towards the center.  This contraction is causing the core size to drop
below our resolution (Fig.~\ref{fig:core}) and the profile slope
$\alpha$ to drop below -1.5 (Fig.~\ref{fig:4pan}).

\begin{figure}
  \includegraphics[width=0.50\textwidth]{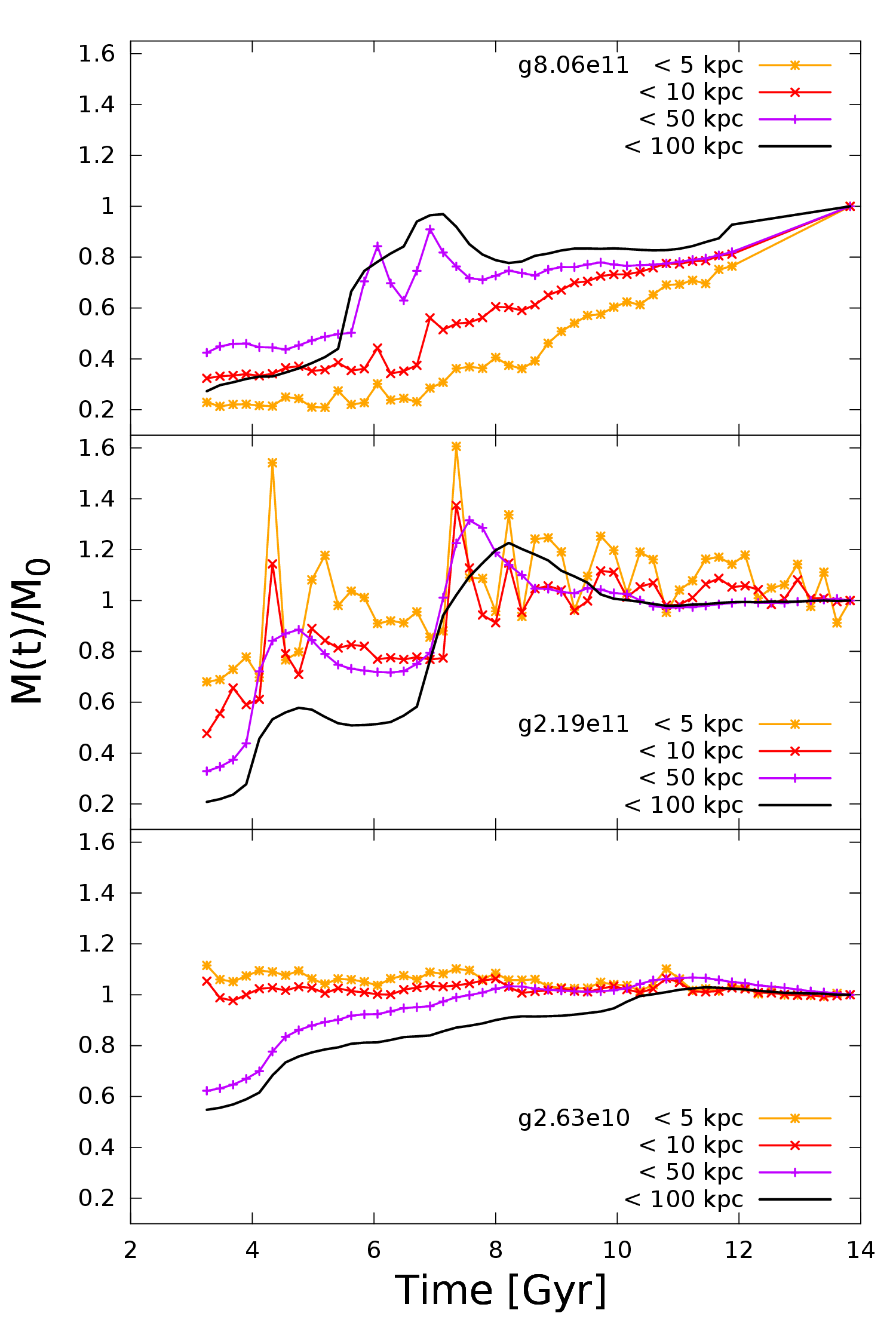}
  \caption{Time evolution of the {\it total} mass enclosed within different radii,
    namely 5,10,50 and 100 kpc, normalized to the $z=0$ mass.
    From top to bottom the results for our three test galaxies: g8.06e11, g2.19e11,
    and g2.63e10. The strong evolution of the inner mass (2 and 5 kpc) for galaxy g8.06e11
    is thought to be the cause of the halo recontraction (see text for more details).}
\label{fig:centralmass}
\end{figure}

The deepening of the central potential (due to gas inflow and star
formation) in our more massive halo can be seen in
Fig.~\ref{fig:centralmass} where we plot the evolution of the {\it
  total} mass (DM+gas+stars) within different radii for our three
galaxies.  In the middle mass halo (g2.19e11) the mass is constant
practically at all radii after 8 Gyrs, this implies a constant
gravitational potential.  In the higher mass halo (g8.06e11) at large
radii (100 kpc, still well within the virial radius) the total mass is
also constant, but at small radii (5 and 10 kpc) the mass rapidly
increases by a factor of three after t=8 Gyrs.  This higher mass
caused the gravitational potential to deepen and reduces the ability
of baryonic feedback to counteract the DM contraction.

It is worth noticing that the change in the mass distribution is mainly
local (i.e. limited to the regiond dominates by baryons) and not global,
something that should be taken into account when
computing analytic estimates of the total energy input from SN
needed to alter the DM distribution. This is a still quite
debated topic in the literature and our simulations can be used to determine
the validity of different analytic approaches (e.g.  Pe{\~n}arrubia \etal 2012 and Maxwell \etal 2015).

\begin{figure}
  \includegraphics[width=0.49\textwidth]{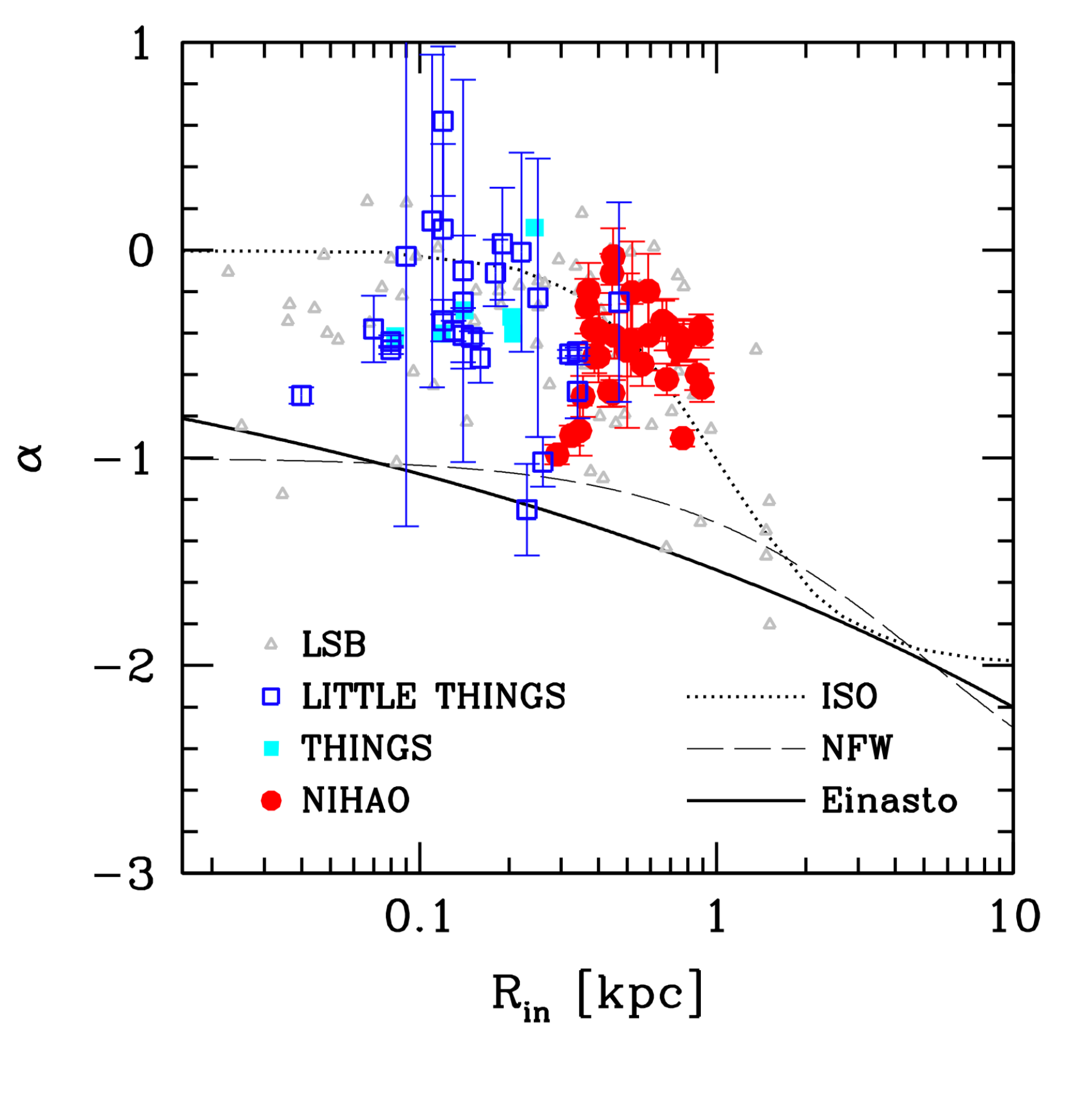}
  \caption{ The inner slope of the dark matter density profile
    $\alpha$ vs.  the radius $R_{\rm in}$ of the innermost point
    within which $\alpha$ is measured.  Grey symbols are results from
    LSB galaxies (de Blok et al. 2001; de Blok \& Bosma 2002; Swaters
    et al. 2003), cyan filled squares are the results from the THINGS
    survey (Oh \etal 2011a), open blue squares with error-bars are the
    results from the LITTLE THINGS survey (Oh \etal 2015).  The red
    points are all NIHAO galaxies with stellar masses between
    $10^7-10^{10} \Msun$ to mimic the  range of the observational
    points. The Isothermal profile line is for a core of 1 kpc, while
    the NFW and Einasto lines are for a concentration of 14 (Dutton \&
    Macci\`o 2014).  }
\label{fig:data}
\end{figure}

\section{Comparison with Observations}\label{sec:obs}

There is a very large amount of literature on the DM density profile
in observed galaxies (e.g Moore 1994, Salucci \& Burkert 2000; Dutton
\etal 2005; Simon \etal 2005; de Blok \etal 2008; Kuzio de Naray,
McGaugh \& de Blok 2008; Kuzio de Naray, McGaugh \& Mihos 2009; Oh
\etal 2011a, Oh \etal 2015).  Despite the quite large scatter in the
slope of the observed profiles  there is a clear indication for
$\alpha_{\rm obs}>-1$ (e.g. Oh \etal 2011a), which is in contrast with
naive expectations from collisionless simulations.  This disagreement
is strongly alleviated when the effects of baryons are taken into
account, as recently shown in Karukes, Salucci \& Gentile (2015),
where the authors compared the measured DM density profile in NGC 3198
with the predictions from the DC14 model.  Thanks to the large number
of simulated galaxies in the NIHAO project we can for the first time
extend previous observation-simulation comparisons that were based on
a limited number of objects (e.g. Oh \etal 2011b).

In Fig.~\ref{fig:data} we show the inner slope of the dark matter
density profile $\alpha$ vs. the radius $R_{\rm in}$ of the innermost
point within which $\alpha$ is measured for both observations and
simulations.  The observational data are split in three groups, the
grey symbols are results from dwarf and Low Surface Brightness (LSB) galaxies
(de Blok et al. 2001; de Blok \& Bosma 2002; Swaters et al. 2003),
cyan filled squares are the results from the THINGS survey (Oh \etal
2011a), while open blue squares with error-bars are the results from
the LITTLE THINGS survey (Oh \etal 2015). The three lines show
theoretical predictions from a given profile: the dotted line is for
an isothermal profile with a core size of 1 kpc, the solid line is for
the Einasto profile (Einasto 1965, Merritt \etal 2005) with a
concentration of 14 as expected for dark matter haloes around $10^{11}
\Msun$ (Dutton \& Macci\`o 2014) and the dashed line is for the NFW
profile with the same concentration as the Einasto one.  It is
interesting to note that on the radial scales of this plot, the
Einasto profile (which is known to provide a better fit to
simulations, Merritt \etal 2005, Navarro \etal 2010) predicts a
steeper slope for DM haloes with respect to NFW.

The red circles with error-bars are the predictions for $\alpha$ from the NIHAO
simulation suite at 1-2\% of \Rvir, for galaxies with stellar masses between
$10^7-10^{10} \Msun$ to be consistent with the LITTLE THINGS sample
range (Oh \etal 2015).  All the NIHAO dark mater haloes are 
expanded with respect to collisionless simulations expectations.
We don't have enough resolution to probe the inner part of the radial profile
at the same extent as the LITTLE THING survey, nevertheless our results
are in very good agreement with the observations where they overlap in
$R_{\rm in}$. Moreover thanks to the large number of simulated
galaxies, we are also able to reproduce the observed scatter in
$\alpha$ at a fixed radius.

This plot further demonstrates the importance of considering the
effect of baryons on the dark matter distribution when comparing
observations and simulations based on the Cold Dark Matter model.

\section{Conclusions}\label{sec:conclusion}

We have used the NIHAO simulation suite (Wang \etal 2015) to study the
effects of baryons on the radial distribution of dark matter in
collapsed haloes.  In comparison with a similar study performed
earlier with the MaGICC sample (Di Cintio \etal 2014a, DC14) we have a
larger and more coherent set of simulations, which have been performed
with the latest compilation of cosmological parameters from the Planck
Collaboration (2014).  Moreover we use a substantially improved
version of the {\sc gasoline} code, which thanks to a new formulation
of the SPH force computation (Keller \etal 2014), aims to solve the
known short comings of the classical SPH implementation (Agertz \etal
2007; Hopkins 2013).

Overall we confirm previous findings from DC14, namely that the effect
of baryons on the slope ($\alpha$) of the central distribution of dark
matter strongly depends on the ratio between the stellar mass and
total mass of the halo.
At the present time, haloes with an intermediate ratio of
stellar-to-halo mass ($10^{-3}<\Mstar / \Mhalo<10^{-2}$ tend to expand
their profiles with respect to pure collisional simulations, haloes
with lower values of $\Mstar / \Mhalo$ tend to retain the original
dark matter cuspy profile, while haloes with large ($\Mstar /
\Mhalo>10^{-2}$) ratio exhibit contracted profiles.  There is a smooth
transition among these different behaviors and we have parameterize it
with a new, simple but accurate fitting formula.
The slopes predicted by the NIHAO simulation suite are in excellent
agrement with high resolution observations of dwarf galaxies
from the THINGS and LITTLE THINGS surveys (Oh \etal 2015).

When the analysis is, for the first time, extended to higher
redshifts, we find that haloes tend to cluster around the redshift
zero $\alpha$ vs. $\Mstar / \Mhalo$ relation, even though the scatter
around the mean increases with redshift.
Our findings imply that even haloes that have very cuspy profiles
today (as for example for Milky Way mass haloes), likely had a cored
dark matter distribution in the past.

We carefully looked at three specific haloes with masses around
$10^{10}, 10^{11}, 10^{12} \Msun$. The lightest shows a central slope
of about $-0.9$ at $z=0$, which is consistent with a cuspy profile,
but is also very much shallower than what a pure N-body simulations
predicts at the same mass and radii ($\alpha \sim -1.6$), pointing to
a clear halo expansion even at this low mass scales.
The density profile was even shallower at higher redshift ($z\sim 1$)
when the halo went through several star formation episodes.

The other two haloes present a quite dynamic evolution in their central
regions as a function of time, showing that not only cores are
created and destroyed during the galaxy evolution but that this
phenomenon happens on relatively (cosmologically) short time scales of
the order of a Gyr.

Overall our results show that one of the most firm predictions of the
Cold Dark Matter theory, namely the existence of a ``universal
profile'', is no longer valid when galaxy formation is taken into
account.  The central part of any dark matter halo has had a much more
tumultuous life than what dissipationless simulations predict.

\section*{Acknowledgments}

It is a pleasure to thank Se-Heon Oh, for sending us his data
in electronic form. AVM, AAD, GSS, CP, TAG, and TB acknowledge 
support from the Sonderforschungsbereich
SFB 881 ``The Milky Way System'' (subprojects A1 and A2) of the German
Research Foundation (DFG).  The simulations were performed on the {\sc
  theo} cluster of the Max-Planck-Institut f\"ur Astronomie and the
{\sc hydra} cluster at the Rechenzentrum in Garching; and the {\sc
  Milky Way}  supercomputer, funded by the Deutsche
Forschungsgemeinschaft (DFG) through Collaborative Research Center
(SFB 881) ``The Milky Way System'' (subproject Z2), hosted and
co-funded by the J\"ulich Supercomputing Center (JSC). We greatly
appreciate the contributions of all these computing allocations.  The
authors acknowledge support from the MPG-CAS through the partnership
program between the MPIA group lead by AVM and the PMO group lead by
XK. LW acknowledges support of the MPG-CAS student programme. 
XK is supported by the NSFC (No.11333008) and the Strategic Priority Research Program the emergence of cosmological structure of the CAS (No. XDB09000000)
%




\label{lastpage}

\end{document}